\documentclass[twocolumn,showpacs,preprintnumbers,amsmath,amssymb,eqsecnum]{revtex4}

\usepackage{graphicx}
\usepackage{dcolumn}
\usepackage{bm}

\begin{document}

\preprint{}

\title{Local disorder and optical properties in V-shaped quantum wires :\\ towards one-dimensional exciton systems}

\author{T. Guillet}
 \altaffiliation[Also at ]{D\'epartement de Physique, Ecole Polytechnique F\'ed\'erale de Lausanne, Switzerland}
 \email{guillet@gps.jussieu.fr}
\author{R. Grousson}
\author{V. Voliotis}
 \altaffiliation[Also at ]{Universit\'e Evry-Val d'Essonne, France}

\affiliation{Groupe de Physique des Solides, CNRS,\\
Universit\'es Pierre et Marie Curie et Denis Diderot, \\
 2 place Jussieu, F-75251 Paris Cedex 05, France }

\author{X.L. Wang}
 \altaffiliation[Also at ]{Department of Electrical Engineering, Yale University, USA}
\author{M. Ogura}

\affiliation{Photonics Research Institute, National Institute of
Advanced Industrial Science and Technology (AIST), Tsukuba Central
2, Tsukuba 305-8568, Japan \\
 and CREST, Japan Science and Technology Corporation (JST), 4-1-8 Honcho, Kawaguchi 332-0012, Japan}

\date{\today}

\begin{abstract}
The exciton localization is studied in GaAs/GaAlAs V-shaped quantum wires (QWRs) by high spatial resolution spectroscopy. Scanning optical imaging of different generations of samples shows that the localization length has been enhanced as the growth techniques were improved. In the best samples, excitons are delocalized in islands of length of the order of $1 \ \mu m$, and form a continuum of 1D~states in each of them, as evidenced by the $\sqrt{T}$ dependence of the radiative lifetime. On the opposite, in the previous generation of QWRs, the localization length is typically $50 \ nm$ and the QWR behaves as a collection of quantum boxes. These localization properties are compared to structural properties and related to the progresses of the growth techniques. The presence of residual disorder is evidenced in the best samples and explained by the separation of electrons and holes due to the large in-built piezo-electric field present in the structure.
\end{abstract}

\maketitle

\section{Introduction}

The local disorder in semiconductor heterostructures is crucial for the understanding of their physical properties. It leads on a macroscopic point of view to two well-known features in optical spectra~: the inhomogeneous broadening of the transitions due to the size and/or composition fluctuations, and the Stokes-shift between photoluminescence and absorption spectra which is a signature of localization. The major role of monolayer fluctuations of the hetero-interfaces has first been identified in quantum wells (QW) \cite{Bastard-PRB84, Yang-PRL93} and is even more important in quantum wires (QWRs) due to the increased number of interfaces involved in the electronic confinement. The study of the local optical properties by high spatial resolution spectroscopy, in far-field \cite{Bellessa-APL97} or in near-field \cite{Intonti-PRB01, Crottini-PRB01}, allowed a better insight into the understanding of localization in these structures. 

In the best QWR structures grown up to now it has been shown that the random potential felt by the excitons and induced by the structural disorder \cite{Zimmermann-PSSA97} completely localizes the excitons~: the localization length is small and each minimum of the potential behaves as a ``natural'' quantum box \cite{Wu-PRB00}, so that the optical properties of the QWR can be interpreted as those of a zero-dimensionnal (0D) system \cite{Bellessa-PRB98, Guillet-PhE01}. The specific characteristics of a one-dimensionnal (1D) system, especially its singular density of states, are therefore blurred out by the disorder. Until now delocalized excitons could only be observed at some given positions of QWR samples as their structural properties were improved \cite{Crottini-PRB01}, opening the way to the achievement of actual 1D~systems.

In this paper we investigate the localization properties of the recent generation of QWRs and compare them to previous samples. The disorder is first studied by our new scanning optical imaging spectroscopy technique. In particular the localization length is measured (Part~\ref{sec:imaging}). Its mean value is shown to be ten times larger than in the previous samples, and is much larger than the thermal de Broglie wavelength of the excitons (Part~\ref{sec:de-Broglie}). These new QWRs can therefore be considered as actual 1D~systems even on a microscopic point of view. This interpretation is supported by the study of the radiative lifetime as a function of temperature, which is highly dependent on the dimensionnality of the local density of states in the structure (Part~\ref{sec:tau-rad}). Finally the delocalized excitons are shown to be more sensitive to the presence of a residual disorder in the structure (Part~\ref{sec:residual-disorder}).

The nanostructures studied experimentally are undoped $5$\ nm thick GaAs/Ga$_{0.57}$Al$_{0.43}$As V-shaped QWRs. They are grown on a $4$\ $\mu$m pitched V-grooves formed on $(001) \pm 0.1^{\circ}$ GaAs substrates by flow rate modulation epitaxy which is a modified metal organic vapor phase epitaxy technique \cite{Wang-JCrystG00A, Wang-JCrystG00B}. Their main features are a very strong lateral confinement leading to energy separation between subbands up to 60\ meV, and a large optical anisotropy, characteristic of the valence band mixing in 1D structures \cite{Wang-APL95}.

\section{Imaging the disorder in a quantum wire}
\label{sec:imaging}

In order to understand exciton localization and its influence on the spectroscopy and the dynamics of QWs or QWRs, we have to model and somehow measure the localization potential felt by the excitons. It is characterized by two main quantities~: its amplitude $V_{loc}$ and its typical correlation length $L_{loc}$. $V_{loc}$ can be related to macroscopic properties like the Stokes shift or the half width of the photoluminescence spectrum, under some assumptions on the shape of the potential \cite{Yang-PRL93}. Its value of $7-10 \ meV$ is quite independent of the sample since it simply reflects the change of the confinement energy associated to monolayer fluctuations on the hetero-interfaces. On the contrary the correlation length $L_{loc}$ is closely related to the roughness of the interfaces and depends much more on the growth conditions. This parameter can only be evaluated through microscopic studies which give access to the spatial extension of the excitonic states.

We recently developped a new imaging technique in order to study the localization effects in quantum wires. This is an evolution of the microphotoluminescence ($\mu$-PL) experiment, that we called scanning optical imaging spectroscopy. It is the equivalent in far-field to the scanning near-field optical microscopy (SNOM). The sample is fixed on the cold finger of a Helium cryostat and cooled at $10 \ K$. The laser beam is focused on the sample by a microscope objective with a large numerical aperture ($0.6$) and the laser spot diameter of $1 \ \mu m$ is diffraction-limited. The luminescence is analyzed through an imaging spectrometer coupled to a liquid nitrogen cooled CCD detector, and the spectral resolution is $50 \ \mu eV$. The excitation at $1.77 \ eV$ creates carriers in the higher order bands of the wire and not in the barriers. As the wires are separated by $4 \ \mu m$, only one single wire is excited over $1 \ \mu m$ in our experiment. The excitation spot can be translated with respect to the sample by moving the microscope objective with piezoelectric actuators with a precision better than $200 \ nm$. At each position the $\mu$-PL spectrum is recorded. We obtain in this way a high resolution image along the wire axis as well as a high resolution spectrum of a single QWR.

\begin{figure}[htb]
\begin{center}
\includegraphics{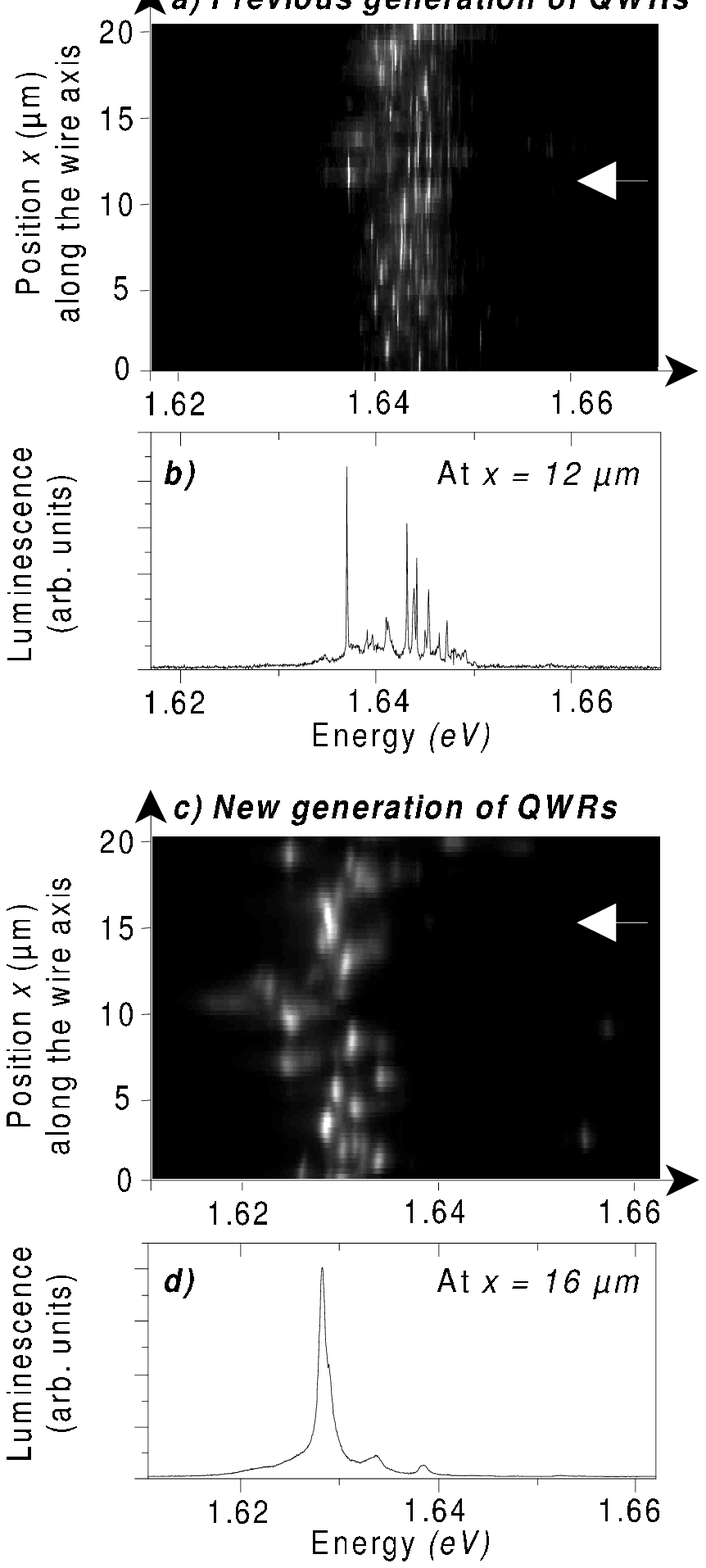}
\end{center}
\caption{Scanning optical images of a single wire of the previous (a) and the new (c) generation of QWRs. The $\mu$-PL intensity is represented in gray levels. Typical $\mu$-PL spectra extracted from these images at the positions indicated by white arrows, are presented in (b) and (d).}
\label{fig:images}
\end{figure}

Figure~\ref{fig:images} presents two such scanning images obtained on two different generations of quantum wires ---denoted previous and new generations hereafter---, and two corresponding $\mu$-PL spectra extracted from the images. The sample of the previous generation studied in this work has already been the subject of $\mu$-PL studies \cite{Bellessa-PRB98}, whereas the new generation is much more recent \cite{Liu-APL01}. As it is well known for a few years \cite{Bellessa-APL97}, the emission spectrum isn't uniform as the excitation spot scans the wire, which means that the excitons aren't delocalized over the whole wire. Each bright spot corresponds to a local potential minimum along the wire axis where excitons are trapped. Each peak has been attributed to the emission of the lowest lying level in such a minimum. Two main differences can be distinguished between the two generations of QWRs~: (i) The spectra associated to each localization sites are different~: the peaks are very sharp for the previous generation, and their linewidth couldn't be resolved within our experimental resolution, whereas they are $1 \ meV$ broad in the case of the new generation. (ii) The linear density of sites, estimated by counting the spots on the images, is of the order of $10 \ \mu m^{-1}$ for the previous generation of QWRs, and $1 \ \mu m^{-1}$ for the new one. We can therefore roughly estimate the extension of the sites to $100 \ nm$ for the previous generation, which is compatible with the value obtained from $\mu$-PLE studies \cite{Bellessa-PRB98}, and to $1 \ \mu m$ for the new generation. Moreover, some emitting sites are extended over distances larger than the spatial resolution of the microscope in the new samples, the spot at $x=15 \ \mu m$ being for example $3 \ \mu m$ long on Fig.~\ref{fig:images}.c. Most of the emitting sites could however not be resolved spatially because their extension is close to or smaller than the resolution.


\begin{figure}[htb]
\begin{center}
\includegraphics{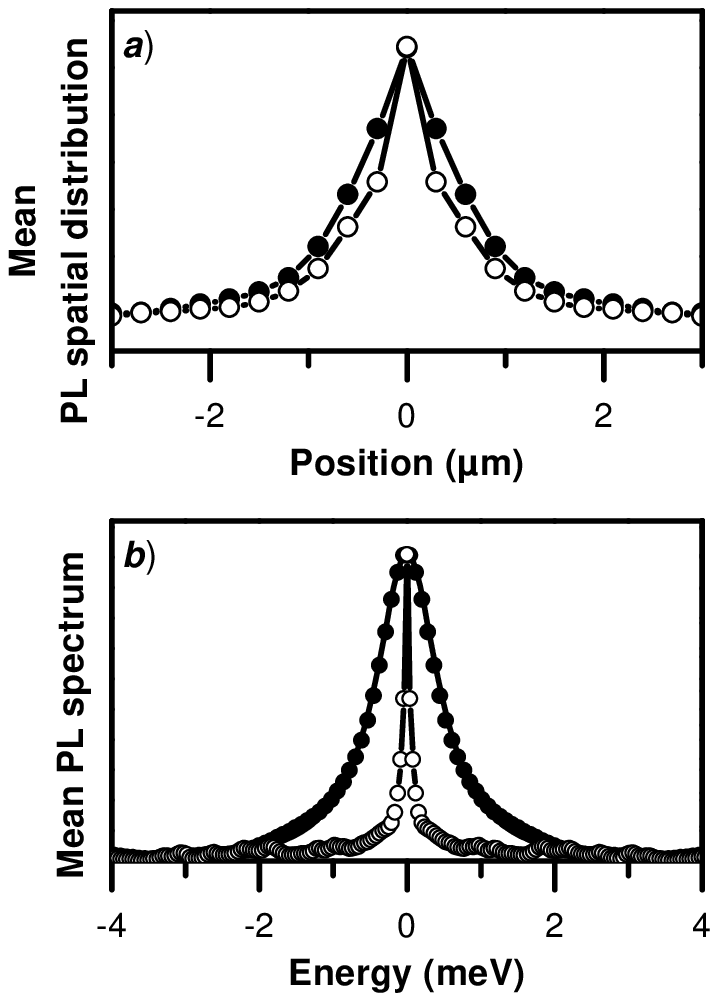}
\end{center}
\caption{Statistical mean properties of the localization sites in the previous (open dots) and new (plain dots) generations of QWRs~: (a) Mean spatial distribution of the emitted luminescence. (b) Mean spectrum; The plain line is a fit by a lorentzian lineshape.}
\label{fig:mean-spectra}
\end{figure}

These properties have been studied and compared more quantitatively through a statistical analysis of the images, which is described in details in the appendix. It allows us to define two mean properties of the localization sites in a given sample~: the mean spectrum and the mean spatial distribution of its luminescence. They are presented on figure~\ref{fig:mean-spectra}. In the spatial dimension (Fig.~\ref{fig:mean-spectra}.a), the mean distribution of the luminescence emitted by one site has a full width at half maximum (FWHM) of $0.8 \ \mu m$ for the previous generation, corresponding to the spatial resolution of the microscope, since the emitting sites are much smaller than the resolution and can be considered as point-sources. The larger FWHM ($1.2 \ \mu m$) obtained for the new generation is due to the extension of the sites, which becomes comparable to the resolution of the microscope and can be estimated to $1.2 \ \mu m - 0.8 \ \mu m = 0.4 \ \mu m$. In the spectral domain (Fig.~\ref{fig:mean-spectra}.b), the mean spectrum of the emitting sites for QWRs of the previous generation is very sharp as seen on individual spectra, whereas the one for the new generation of QWRs is perfectly fit by a $1 \ meV$ broad lorentzian line, which is even more clearly evidenced by the perfect triangular shape of its Fourier transform on a logaritmic scale over 4 decades, shown in the figure~\ref{fig:autocorrelation}.c of the appendix. This broadening will be discussed in more details in part~\ref{sec:residual-disorder}.


The samples of the new generation are therefore of very high quality. For comparison, the longest reported localization length in V-groove QWRs as well as in any other kind of GaAs quantum wires is of the order of $600 \ nm$ and has been measured by SNOM only at one position of a given sample \cite{Crottini-PRB01}; the typical extensions of the localization sites in the sample studied in this work were of 10 to $50 \ nm$, and are comparable to those observed by our technique in the samples of the previous generation. In the new generation of QWRs, the large localization length of $400 \ nm$ is a mean property of the whole sample and obtained after a statistical analysis, the largest observed emitting sites being 2 to $3 \ \mu m$ long. The consequences of this improvement on the optical properties of the excitons are discussed in the next part.

\begin{figure}[htb]
\begin{center}
\includegraphics{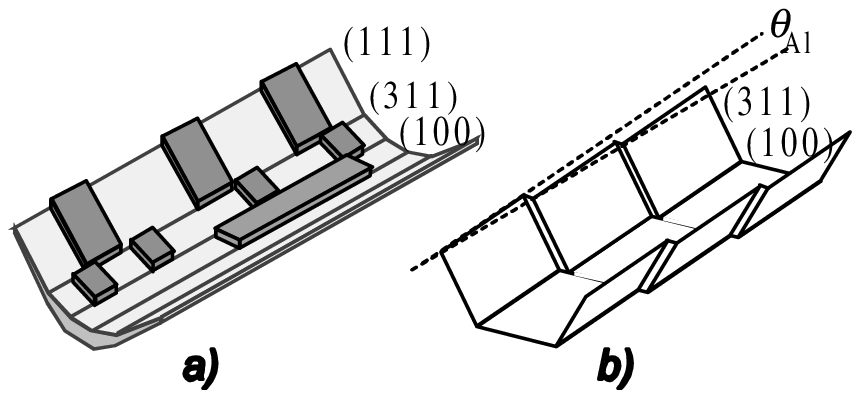}
\end{center}
\caption{(a) Monolayer fluctuations of the interfaces between the well and the barrier materials of the V-shaped wire; (b) Monolayer steps induced by a misalignment of the V-groove and the crystallographic direction $[1\bar{1}0]$.}
\label{fig:schema-ML}
\end{figure}

The enhancement of the localization length of excitons in the wires has been related to the structural properties of the samples. Indeed the localization of excitons is mainly due to the monolayer fluctuations of the interfaces of the wire, as schematically represented on figure~\ref{fig:schema-ML}.a. These interfaces have been identified on transmission electron micrographs (TEM) and are locally flat at the atomic level \cite{Wang-JCrystG00A}, but they present monolayer steps which modify the confinement energy of the excitons in the wire. We calculated that the energy fluctuation associated to a monolayer step on the central facet (100) is about $9 \ meV$, and the one for the lateral facets (311) and (111) is equal to $3 \ meV$. These fluctuations can be directly identified on the images obtained on QWRs of the new generation (Fig.~\ref{fig:images}.c)~: the $3 \ meV$ ones are the most frequent (1 to 2 per $\mu m$) whereas the larger ones (about $10 \ meV$, like at $x=10 \ \mu m$) are fewer (1 every $5 \ \mu m$). These values correspond to the linear density of monolayer steps on the lateral and central facets respectively, typically one step every $500 \ nm$. 


Three main changes in the fabrication of the samples have been developped in the past years. First of all, the chemical etching of the patterned sample allowed to get smoother hetero-interfaces between the well and the barrier materials, as shown by atomic force microscopy (AFM) images \cite{Wang-JCrystG00A}. Second, the arsenic source for the epitaxy has been changed and tertiarybutylarsine ($TBAs$) is now used instead of the very toxic arsine ($AsH_3$). This leads to a strong diminution of the non-radiative recombination at room-temperature, which is interpreted as the reduction of the impurity concentration in the sample \cite{Wang-JCrystG00B, Liu-APL01}. The last and most crucial parameter is the misalignment angle $\theta_{Al}$ between the V-groove drawn by lithography and the crystallographic direction $[1\bar{1}0]$ (Fig.~\ref{fig:schema-ML}.b). We performed scanning optical imaging spectroscopy on samples grown in the same conditions but presenting different misalignements. The statistical analysis revealed that the linear density of emitting sites decreases from $4 \ \mu m^{-1}$ to $1 \ \mu m^{-1}$ as the angle $\theta_{Al}$ is decreased from $0.1^{\circ}$ to less than $0.004^{\circ}$ by improving the alignment procedure. This confirms that the localization length is governed in good samples by the distance between monolayer steps on the lateral interfaces of the V-groove, {\it i.e.} by the alignment of the V on the crystallographic direction.

\section{Localization, excitonic eigenstates and density of states}
\label{sec:de-Broglie}

The enhancement of the localization length in the wires implies not only a quantitative, but also a qualitative change of their electronic properties. First of all, let us briefly consider the competition between the electrons and holes localization and the Coulomb interaction. In both generations of QWRs, the amplitude of about $10 \ meV$ of the localization potential is smaller than the exciton binding energy, which is of the order of $20 \ meV$, and the localization length scale is larger than the exciton Bohr radius $a_X = 7 \ nm$ \cite{Guillet-X1D}. The localization hamiltonian can therefore be treated as a perturbation on the free rigid exciton model, and the localization potential felt by the excitons is simply the average over the wavefunction of the exciton relative motion ({\it i.e.} over $a_X$) of the potential felt by electrons and holes.

\begin{figure}[htb]
\begin{center}
\includegraphics{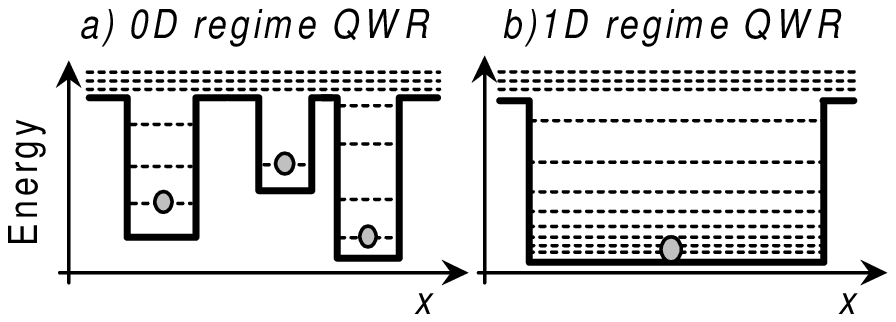}
\end{center}
\caption{Schematic representation of the eigenstates of the localization potential in the previous generation of QWRs, in the 0D~regime ($V_{loc}=9 \ meV$, $L_{loc}\approx 30 \ nm$) (a) and in the new generation, in the 1D~regime ($V_{loc}=9 \ meV$, $L_{loc}\approx 500 \ nm$) (b).}
\label{fig:regimes}
\end{figure}

We should also compare the localization length $L_{loc}$ of the excitons to their thermal de Broglie wavelength  $\lambda_{Th} = \sqrt{\frac{2 \pi \hbar^2}{m_X k_B T}}$, where the exciton mass $m_X \approx 0.18 \ m_0$ is the sum of the electron and hole masses \cite{masses}. It is equal to about $50 \ nm$ at $T=10 \ K$ in the wires we have studied. If $L_{loc} > \! > \lambda_{Th}$, the energy difference between the eigenstates of the localization potential, typically $\frac{\hbar^2 \pi^2}{2 m_X L_{loc}^2}$, is much smaller than the thermal energy $k_B T$ and these eigenstates can be approximated by a 1D-continuum of excitons. In the opposite case, when $\lambda_{Th}$ is comparable to or larger than $L_{loc}$, only one or two localized states are populated at thermal equilibrium and the density of localized states is discrete. Since most of the microscopic properties of the excitons ---relaxation, recombination--- are determined by the local exciton density of states, the criterium $\lambda_{Th} < \! < L_{loc}$ determines the one-dimensional character of the system. In the quantum wires of the previous generation, it isn't fulfilled and the optical properties are those of a collection of quantum boxes, as our previous studies of the exciton relaxation, recombination \cite{Bellessa-PRB98} (cf Part.~\ref{sec:tau-rad}) and fine structure \cite{Guillet-PhE01} have demonstrated; these wires are therefore said to be in the ``0D~regime''. On the contrary, in the new generation of QWRs $\lambda_{Th} < \! < L_{loc}$, so that each localizing site can be considered as a small portion of quantum wire with a continuum of 1D states; these sites are called ``islands'' in the following, in analogy with the situation in quantum wells, and the QWRs are told to be in the ``1D~regime''. The two regimes are schematically depicted in the figure~\ref{fig:regimes}.


This interpretation is valid on a local point of view, as far as microscopic properties of the wires are concerned. On a macroscopic scale, the electronic properties are in both regimes inhomogeneously broadened of about 5 to $10 \ meV$ due to the variations of the confinement energy associated to monolayer fluctuations. This is different from the situation in quantum wells, where only one or two sharp peaks associated to two well thicknesses of $N$ and $N+1$ monolayers are observed in the best samples \cite{Deveaud-APL87}. In the case of QWRs, monolayer fluctuations on each of the hetero-interfaces of the V give rise to a large number of possible configurations for the confinement, and therefore to an inhomogeneous broadening of the photoluminescence spectra.

It is here interesting to compare the different techniques used to study localization in QWs and QWRs, and to understand how they are adapted to specific localization regimes. In this paper we present two methods~: the scanning imaging of the sample allows to directly measure the localization length in 1D~regime QWRs, when $L_{loc}$ is comparable to the resolution of the microscope. But in 0D~regime QWRs $L_{loc}$ is too small and only the linear density $n$ of localizing sites can be obtained from the images, which gives an upper limit $L_{loc} < 1/n$. In 0D~regime QWRs, we have shown in previous works \cite{Bellessa-PRB98} that the PLE spectra of each quantum box allow to determine their extension from the energy separation between ground and excited states. But this is only valid if the quantum boxes are independent and if they contain two or more confined states, {\it i.e.} if $V_{loc} > \hbar^2 \pi^2 / (2 m_X L_{loc}^2)$. For smaller amplitude or typical length of the potential, the localized excitonic states extend over a few potential minima if $V_{loc} < \hbar^2 \pi^2 / (2 m_X L_{loc}^2)$, and the eigenenergies are strongly spatially correlated. This is known as the level repulsion. Recently Idrissi et al. reported evidences for a level repulsion in QWRs very similar to our 0D~regime QWRs \cite{Idrissi-02}. In our statistical analysis, we were able to extract the same autocorrelation function ($R_c(E)$) as they do, but we didn't see any signature of this effect in our results. This can be explained by two practical reasons~: (i) the cusp in the $R_c(E)$ which is characteristic of the level repulsion may be too small to be distinguished from the statistical noise in our case, even if we averaged over a $35 \ \mu m$ long image; (ii) our spatial resolution is four times larger than the one of the SNOM experiment used in this work and is therefore much larger than the localization length, so that the spatial filtering of the correlations is less efficient. But it could also reflect that our 0D~regime QWRs fulfill the condition $V_{loc} > \hbar^2 \pi^2 / (2 m_X L_{loc}^2)$, so that the quantum boxes can be considered as isolated and no level repulsion exists if only the ground states emit luminescence. This is supported by our full understanding of the dynamics and spectroscopy of the excitons in 0D~regime QWRs \cite{Bellessa-PRB98} as those of isolated quantum boxes.

\section{Radiative recombination and localization regime}
\label{sec:tau-rad}

The optical property which depends at the most on the density of states is the radiative lifetime $\tau_{rad}$ \cite{Feldmann-PRL87, Citrin-PRL92, Akiyama-PRL94}. Its dependance on the temperature directly reflects the dimensionality of the exciton system~: $\tau_{rad}$ scales as $T$ for quantum wells, $\sqrt{T}$ for perfect quantum wires, and is independent of temperature for quantum boxes as soon as $k_B T$ is smaller than the energy separation between the two first states of the quantum box.

In 0D~regime QWRs, the radiative lifetime has already been studied in $\mu$-PL in previous works \cite{Bellessa-PRB98}. It is strongly dependent on the quantum box under study and is in fact inversely proportional to the length of the quantum box. It is independent of temperature below $20 \ K$. These results were interpreted in the following way~: the typical energy separation between the quantum box states, of 2 to $5 \ meV$, is larger than $k_B T$, so that at $T=10 \ K$, only the ground state of the quantum box is populated and $\tau_{rad}$ is independent of $T$. The oscillator strength of the ground state is proportional to the ratio of the coherence volume to the volume of the exciton relative motion wavefunction \cite{Rashba-SPSS62, Feldmann-PRL87}. The latter one is independent of the localisation length since $L_{loc} > \! > a_X$, and the coherence volume is proportional to the length $L_{loc}$ of the quantum box. So that the radiative lifetime at $10 \ K$ is inversely proportional to the length of the quantum box, as it has been measured experimentally \cite{Bellessa-PRB98}.

\begin{figure}[htb]
\begin{center}
\includegraphics{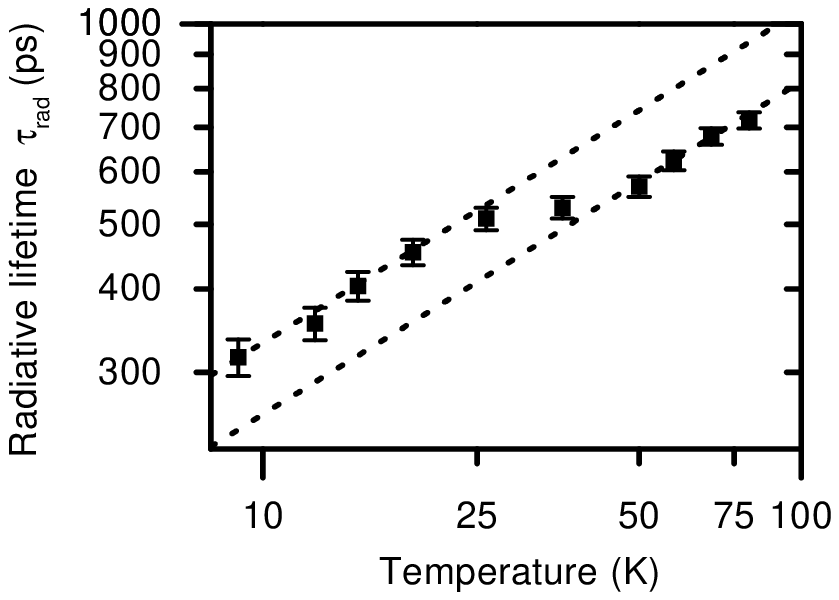}
\end{center}
\caption{Temperature dependence of the radiative lifetime of a single extended island in a 1D~regime QWR. Two fits by a $\sqrt{T}$ dependence are presented in dotted lines (see text).}
\label{fig:tau-rad}
\end{figure}

In 1D~regime quantum wires, the radiative lifetime has been measured on several different wires and several samples, and is at $10 \ K$ independent of the island under consideration within $10 \%$. The results obtained on one island as a function of temperature are presented on figure~\ref{fig:tau-rad}. The island chosen in the presented results is extended over $2 \ \mu m$ according to the scanning optical images. The measurements are performed in the low density regime, and in average one exciton at the most is present in the excited island. The decay time of the PL is equal to the radiative lifetime since non-radiative recombination enters into play only above $200 \ K$ in these samples \cite{Liu-APL01}. $\tau_{rad}$ follows a $\sqrt{T}$ scaling law below $30 \ K$ and above $50 \ K$, which is characteristic of a 1D system. We don't observe any saturation of $\tau_{rad}$ at low temperature down to $8 \ K$, which is the lowest temperature we can achieve in our $\mu$-PL cryostat.


The $\sqrt{T}$ dependence of $\tau_{rad}$ can be briefly explained in the following way~: if we consider free excitons, only those with a wavevector smaller than the one of light $k_\nu=2 \pi / \lambda$ can recombine radiatively, where $\lambda$ is the wavelength of light in the material. The radiative exciton states form a band which spectral width is given by $\Delta_{rad} = \hbar^2 k_{\nu}^2 /2m_X$, and are assumed to have all the same recombination time $\tau_{rad}^0$. At low density, the excitons in the band follow a Boltzman distribution and the effective radiative lifetime of the wire is given by the thermal equilibrium between radiative and non-radiative states~: if $\Delta_{rad} < \! < k_B T$ \cite{Feldmann-PRL87, Citrin-PRL92},
\begin{equation}
\tau_{rad} = \tau_{rad}^0 \sqrt{\frac{\pi \ k_B T}{4 \Delta_{rad}}}.
\end{equation}
The intrinsic recombination time deduced from the experimental results below $30 \ K$ is $\tau_{rad}^0 = 170 \pm 20 \ ps$. It is comparable to the value of $150 \ ps$ theoretically predicted by Citrin in such wires \cite{Citrin-PRL92}. We note that it is much larger than in QWs (a few $10 \ ps$), as expected~: for similar sizes, the 1D and 2D~intrinsic lifetime should differ by a factor $\lambda / a_X$ \cite{Akiyama-PRL94}. Moreover, the $\sqrt{T}$ dependence of $\tau_{rad}$ shows that the excitons are locally at thermal equilibrium at the bottom of the band in each island, and their density of states averaged over $k_B T$ scales like $1/\sqrt{E-E_0}$.

The transition around $40 \ K$ isn't fully understood at the moment, but two assumptions can be put forward~: this temperature corresponds to the typical energy fluctuations along the wire axis ($3 \ meV$), so that the exciton density of states could be increased above $3 \ meV$ leading to a diminution of $\tau_{rad}$ above $40 \ K$ with respect to the situation in a single isolated island; the second possible hypothesis is that the coupling between excitons and acoustic phonons becomes strong above $50 \ K$ and could modify the exciton effective mass and/or their coherence time \cite{These}.

Our results are comparable to those obtained by Akiyama et al. \cite{Akiyama-PRL94}, but they are more precise in the low temperature range where localization effects influence the dependence of $\tau_{rad}$ on $T$. Moreover they extend over a wider range of temperature and are not limited above $50 \ K$ by non-radiative recombination processes. The effect of localization in 0D~regime QWRs has also been clearly evidenced by Oberli et al., who observed and modelled the saturation of $\tau_{rad}$ below $20 \ K$ \cite{Oberli-PRB99}.

\section{Residual disorder in 1D-regime quantum wires}
\label{sec:residual-disorder}

The $\mu$-PL spectra of single islands present a broadening of about $1 \ meV$ which has to be explained. The individual spectra and even more clearly the mean spectrum obtained by the statistical analysis (Fig.~\ref{fig:mean-spectra}) exhibit a perfectly lorentzian lineshape which is typical of an homogeneous broadening of the transition. However, the measured linewidth corresponds to a dephasing time $T_2$ for the excitons of the order of a picosecond. This dephasing time is too short to be associated to intrinsic dephasing mechanisms~: we measured a recombination time of the order of $300 \ ps$ and we calculated that the interaction with acoustical phonons leads to a dephasing time of $10 \ ps$ at $T= 10 \ K$ for the excitons at the bottom of the band \cite{These}.

\begin{figure}[htb]
\begin{center}
\includegraphics{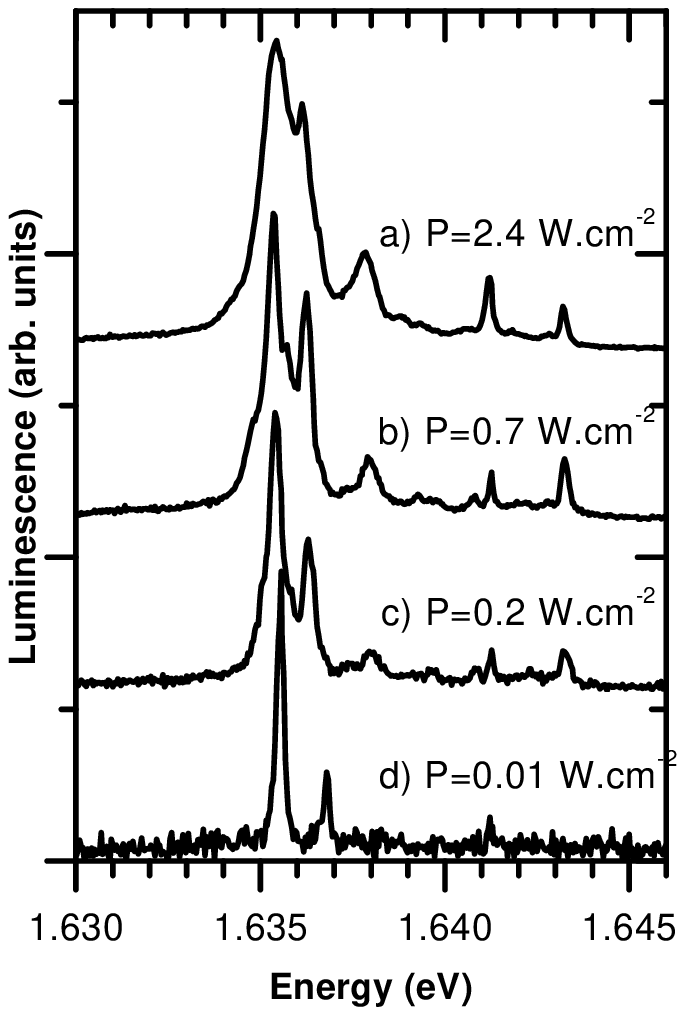}
\end{center}
\caption{PL spectrum of a single extended island in a 1D~regime QWR at very low excitation power.}
\label{fig:residual}
\end{figure}

We therefore have to consider other processes to explain this broadening. A static disorder, such as the alloy disorder in the barrier material or the atomic roughness of the interfaces, would lead to a gaussian inhomogeneous broadening of the transition. But a dynamical disorder, like Coulomb collisions with impurities or charges could be involved and explain the fast dephasing of excitons. A signature of this disorder has been found in the $\mu$-PL spectra under very weak excitation. Figure~\ref{fig:residual} presents the spectrum of a single island extended over $2 \ \mu m$ in a 1D~regime QWR, as a function of the excitation power. We should underline that the excitation power $P_0$ necessary to create one exciton in the excitation spot every $300 \ ps$ (the radiative lifetime at $T= 10 \ K$) is equal to $400 \ W.cm^{-2}$, as it has been measured in 0D and 1D regime QWRs by studying the formation of a biexciton \cite{These, jumeau}. It is 1000 times larger than the excitation power range presented on figure~\ref{fig:residual}. We observe that the spectrum is composed of a main line, corresponding to the emission of the island, and 4 side peaks associated to neighboring islands present in the tails of the excitation spot. The main peak gets narrower as the excitation power is decreased, and it isn't spectrally resolved at the lowest power --- our resolution is $100 \ \mu eV$ in this experiment. This narrowing is therefore occuring as one exciton is created every $500 \ ns$ in the island, which is a crude estimation taking into account the power ratio between $P_0$ and the power range of the experiments presented in figure~\ref{fig:residual}. The integrated intensity of the spectra is proportional to the excitation power, so that there is no new non-radiative recombination channel in this regime. Such line narrowing can only be explained if the electron and the hole are somehow spatially separated in the island since the recombination probability is proportional to the overlap of their wavefunctions.


\begin{figure}[htb]
\begin{center}
\includegraphics{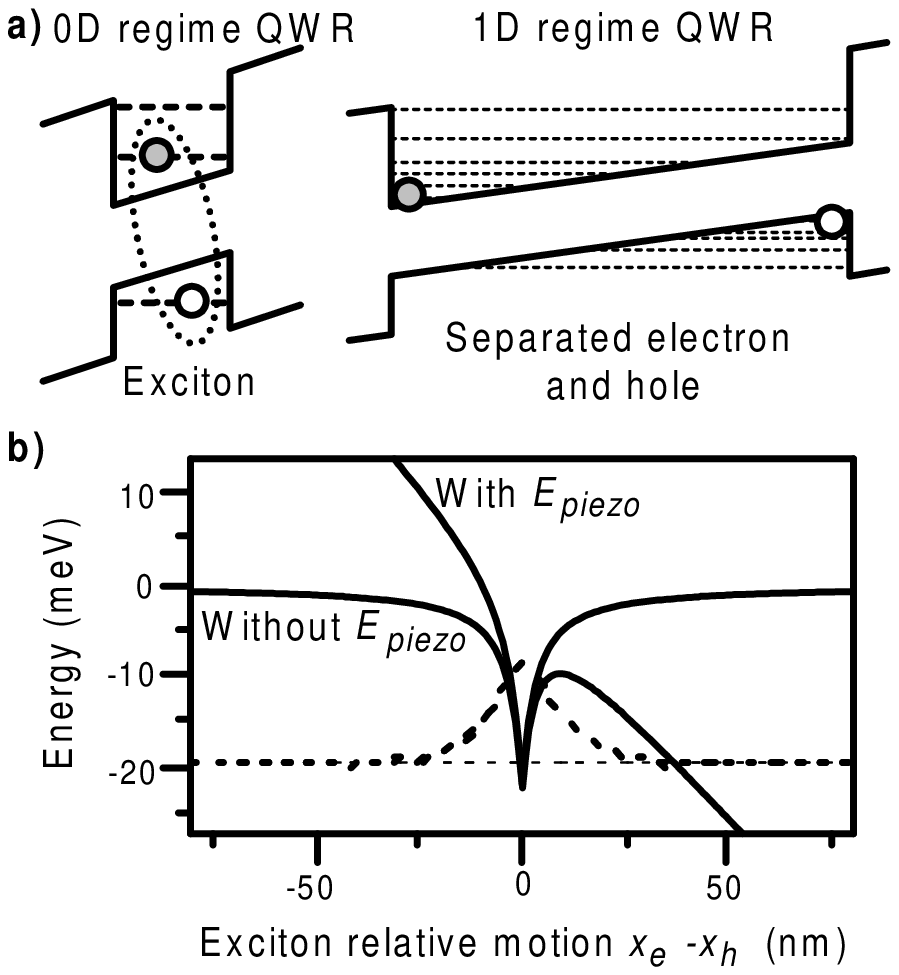}
\end{center}
\caption{(a) Effect of the piezo-electric field along the axis of the wire in both localization regimes; (b) In the case of 1D~regime QWRs, the plain lines represent the potential felt by the relative motion of the electron-hole pair without the piezo-electric field (only Coulomb attraction) and with it (Coulomb + constant electric field), in the referential of the center of motion; the wavefunction of the unperturbed ground exciton state is also shown in dotted line.}
\label{fig:piezo}
\end{figure}

This charge separation is attributed to the large piezo-electric field which exists along the axis of the wire as discussed in Ref.~\cite{Liu-APL02}. The field is sufficient to separate electrons and holes only in 1D~regime QWRs, in which the islands are large enough, whereas in 0D~regime QWRs the exciton confined in a given quantum box is only deformed by the field. This is schematically depicted on figure~\ref{fig:piezo}.a. However the piezo-electric field isn't large enough to break an exciton if it is already formed~: the bound electron-hole state still exists. There are therefore two possible states~: one bound exciton state centered on $x_e-x_h=0$, and one separate electron-hole pair state for large electron-hole separations ($x_e-x_h>50 \ nm$), as shown on figure~\ref{fig:piezo}.b.


We can now explain the evolution of the spectrum with the excitation power~: at the lowest power ($0.01 \ W.cm^{-2}$), only one electron-hole pair is present in the island in a separated state. It forms a bound exciton after a few hundreds of $ns$, which is roughly the transition rate between the separated electron-hole pair state and the bound exciton state. Then it recombines, leading to a sharp PL peak. As the excitation power is increased, and during the linear regime until $P_0$ is reached, a few separated pairs accumulate in the island before they can form excitons, creating a background of free carriers that scatter the excitons, and leading to the broadening of the PL exciton line.

The homogeneous broadening of $1 \ meV$ of the $\mu$-PL peaks in the linear regime (below $P_0$) is herefore a signature of a residual disorder, {\it i.e.} the Coulomb collisions between the excitons and separated carriers. The delocalized character of excitons in 1D~regime QWRs makes them more sensitive to the piezo-electric field. This explain why the broadening was not observed in the spectra of quantum boxes in 0D~regime QWRs in the same range of excitation power.

\section{Conclusion}

By performing scanning optical imaging spectroscopy on different generations of QWRs, we studied the disorder and especially the localization length of the excitons. In the best samples, excitons are delocalized over $400 \ nm$ in average and up to $2-3 \ \mu m$ in the most extended islands. For the first time, QWRs exhibit a localization length much larger than the thermal de Broglie wavelength of the excitons, which therefore form in each island a 1D~quasi-continuum. On the opposite, in previous samples, the localization length is smaller than the de Broglie wavelength and the optical properties of the QWRs are those of a 0D system, {\it i.e.} a collection of quantum boxes. The 1D~character of the local density of states in the best samples is evidenced by measuring the temperature dependence of the radiative lifetime : it follows a $\sqrt{T}$ scaling law down to the lowest achieved temperature ($8 \ K$), whereas a saturation below $20 \ K$ has been observed in previous works on 0D~regime QWRs. Finally we show that the excitons are more sensitive to the residual disorder in the structure as they get delocalized, and the presence of photo-created charges separated by the in-built piezo-electric field leads to an homogeneous broadening of the transitions in $\mu$-PL spectra.

\appendix*
\section{Statistical analysis of the images}

The scanning images of QWRs provide us with an intuitive representation of the excitonic states along the wire, but they also contain more quantitative informations. It is for example very useful to obtain the spatial and spectral shape of the bright spots of the image, {\it i.e.} the spatial distribution and the spectrum of the localizing sites. However this can't be done in a proper way by extracting the cross-sections of one given bright spot because they all slightly overlap, as it can be seen on Figure~\ref{fig:images}. Moreover such cross-sections are specific to the localizing site considered and aren't representative of all of them. We therefore analyzed the images by auto-correlation techniques, and calculated the mean spatial distribution and the mean spectrum of the localizing sites.

In order to present the different calculated correlations, we will suppose for simplicity that an image is composed of $N$ spots with identical shapes but different amplitudes $A_n$, at an energy $E_n$ and a position $x_n$. The $\mu$-PL signal $F(E,x)$ at the position $x$ and the energy $E$ can be written~:
\begin{eqnarray}
F(E,x) &=& \sum_n A_n u_E(E-E_n) \times u_x(x-x_n) \nonumber \\
&=& (u_E(E) \times u_x(x)) \otimes D(E,x),
\end{eqnarray}
where $u_E$ and $u_x$ are the spectral and spatial shapes of the spots, and $D(E,x) = \sum_n A_n \delta (E-E_n) \delta (x-x_n)$ is the distribution of the spectral and spatial positions of the spots. Their Fourier transforms are noted $\tilde{u}_E(s)$, $\tilde{u}_x(k)$ and $\tilde{D}(s,k)$ respectively. Those three functions can be obtained by calculating the auto-convolutions of the image $F$. We will here focus on the spectral analysis, and the same study can be done in the spatial dimension.

\begin{figure}[htb]
\begin{center}
\includegraphics{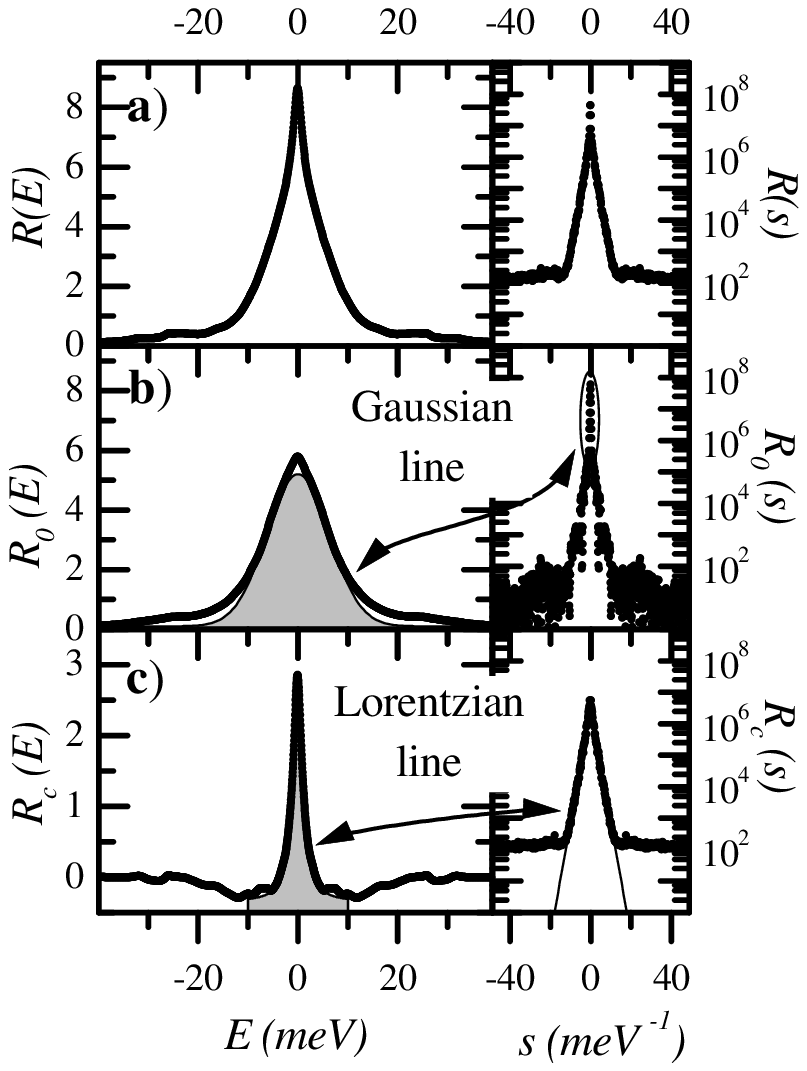}
\end{center}
\caption{(a) Spectral auto-convolution $R(E)$, (b) its spatially uncorrelated component $R_0 (E)$ and (c) the auto-correlation $R_c(E) = R(E)-R_0(E)$, in the case of the image~\ref{fig:images}.c of a QWR of the new generation. The fitted lineshapes (b and c) are presented in filled areas.}
\label{fig:autocorrelation}
\end{figure}

The spectral auto-convolution is defined as~:
\begin{equation}
R(E) = \left\langle \langle F(E', x) F(E-E',x) \rangle_{E'} \right \rangle_x,
\end{equation}
where $\langle \rangle_\alpha$ denotes the average over the parameter $\alpha$. It is the spatial average of the auto-convolution calculated at each position. It is represented on Figure~\ref{fig:autocorrelation}.a with its Fourier transform $\tilde{R}(s) = |\tilde{F}(s,k=0)|^2$, for the image~\ref{fig:images}.c of a QWR of the new generation. It is the sum of two terms~: a spatially uncorrelated part $R_0(E)$ and a correlated one $R_c(E)$.


The uncorrelated auto-convolution (Fig.~\ref{fig:autocorrelation}.b) is equal to~:
\begin{equation}
R_0(E) = \left\langle \langle F(E', x) \rangle_x \langle F(E-E',x) \rangle_x \right \rangle_{E'}.
\end{equation}
Its Fourier transform writes $\tilde{R}_0(s) = \left\langle |\tilde{F}(s,k)|^2 \right\rangle_k$. For a sufficiently large image ($N \rightarrow \infty$), it tends to the auto-convolution of the macro-luminescence spectrum $\langle D(E,x) \rangle_x = \frac{1}{N} \sum_n A_n \delta(E-E_n).$ Its linewidth of $12 \ meV$ corresponds to the inhomogeneous broadening of the PL spectrum multiplied by $\sqrt{2}$, as expected for a gaussian shape.

The correlated part of $R(E)$, called the auto-correlation (Fig.~\ref{fig:autocorrelation}.c), is obtained as~:
\begin{equation}
R_c(E) = R(E) - R_0(E).
\end{equation}
In the limit of large images and if the energy $E_n$ of the spots are spatially uncorrelated, it tends to the auto-convolution of the spectral shape $u_E(E)$ of the spots. In the case of the image~\ref{fig:images}.c, the auto-correlation is mainly composed of a $2.2 \ meV$ broad lorentzian line corresponding to the auto-convolution of $u_E$ and a few side peaks at $\pm 9 \ meV$, $\pm 12 \ meV$, ..., which reflect the correlations between the energies of neighboring islands, {\it i.e.} the energy fluctuations associated to monolayer fluctuations.

In order to evaluate $u_E(E)$, we calculated the ``mean'' spectrum $f_c(E)$ of the islands, represented on figure~\ref{fig:mean-spectra}, as the inverse Fourier transform of $\tilde{f}_c(s) = \sqrt{|\tilde{R_c}(s)|}$. It is correct if we suppose that $u_E$ is an even function of $E$, so that $\tilde{f}_c(s)$ is real. In the case considered in Fig.~\ref{fig:autocorrelation}.c, the lorentzian character of the mean spectrum is even clearer on the Fourier transform $\tilde{R_c}(s)$, which is triangular over 4 decades in a logaritmic scale.

The same analysis can be reproduced in the spatial dimension and provides us with the mean spatial distribution of the luminescence emitted by the islands.


\begin{thebibliography}{100}
\bibitem{masses} We take $m_e = 0.067 \ m_0$ for the electron mass and $m_h = 0.11 \ m_0$ for the heavy hole mass along the wire axis $x$, expressed in terms of the bare electron mass $m_0$.
\end{thebibliography}
\end{document}